\documentclass[aps,prl,twocolumn,superscriptaddress]{revtex4-2}

\usepackage{amsmath,amssymb}
\usepackage{graphicx}

\usepackage[dvipsnames]{xcolor}
\usepackage{amssymb, amsmath}
\usepackage[colorlinks,urlcolor=blue,citecolor=blue,linkcolor=blue]{hyperref}
\usepackage{amsmath}
\usepackage{float}
\usepackage[utf8]{inputenc}
\usepackage{braket}

\begin{document}

\title{Observation of low-lying impurity states in Bose-Einstein condensates}

\author{A. \ M. \ Morgen}
\affiliation{Center for Complex Quantum Systems, Department of Physics and Astronomy, Aarhus University, Ny Munkegade 120, DK-8000 Aarhus C, Denmark}
\author{S.\ S.\ Balling}
\affiliation{Center for Complex Quantum Systems, Department of Physics and Astronomy, Aarhus University, Ny Munkegade 120, DK-8000 Aarhus C, Denmark}
\author{M.\ T.\ Strøe}
\affiliation{Center for Complex Quantum Systems, Department of Physics and Astronomy, Aarhus University, Ny Munkegade 120, DK-8000 Aarhus C, Denmark}
\author{T.\ G.\ Skov}
\affiliation{Center for Complex Quantum Systems, Department of Physics and Astronomy, Aarhus University, Ny Munkegade 120, DK-8000 Aarhus C, Denmark}
\author{M.\ R.\ Skou}
\affiliation{Center for Complex Quantum Systems, Department of Physics and Astronomy, Aarhus University, Ny Munkegade 120, DK-8000 Aarhus C, Denmark}
\author{K.\ K.\ Nielsen}
\affiliation{Niels Bohr Institute, University of Copenhagen, Jagtvej 128, DK-2200 Copenhagen, Denmark}
\author{A.\ Camacho-Guardian}
\affiliation{Instituto de Física, Universidad Nacional Autónoma de México, Apartado Postal 20-364, Ciudad de México C.P. 01000, Mexico}
\author{G.\ M.\ Bruun}
\affiliation{Center for Complex Quantum Systems, Department of Physics and Astronomy, Aarhus University, Ny Munkegade 120, DK-8000 Aarhus C, Denmark}
\author{J. J. Arlt}
\affiliation{Center for Complex Quantum Systems, Department of Physics and Astronomy, Aarhus University, Ny Munkegade 120, DK-8000 Aarhus C, Denmark}
\date{\today}

\begin{abstract} 
Impurities embedded in a Bose-Einstein Condensate (BEC) of $^{39}\text{K}$ atoms are investigated with a pump-probe ejection spectroscopy sequence. The spectroscopic signal exhibits a strong feature corresponding to a Bose polaron in agreement with prior injection spectroscopy and theory. In addition, significant spectral weight at energies well below the energy of the polaron is observed, which is absent in injection spectroscopy. The energy and spectral weight of this signal are measured as a function of interaction strength and evolution time between the pump and probe pulses. We tentatively compare these results to two different theoretical models: a low-energy impurity state dressed by many bosonic excitations and a bipolaron state formed by two polarons due to attractive interactions mediated by the BEC. Such states can exist due to the large compressibility of the weakly interacting BEC. Both theories predict ejection spectra consistent with the low-energy signal, but only the bipolaron model is compatible with its spectral weight. These results indicate that low-energy states below the usual polaron exist for strong interactions, calling for further experimental investigations. 
\end{abstract}

\maketitle
\paragraph{Introduction--} Ultracold quantum gases offer an outstanding platform for investigating strongly interacting quantum many-body systems. One successful application is the immersion of impurity atoms in such gases, which leads to the realization of highly tunable quasiparticles called polarons, consisting of the impurity dressed by excitations in the surrounding gas~\cite{massignanrev2025}. Such experiments have significantly improved our understanding of quasiparticles, which offer a simple yet powerful framework to describe interacting quantum many-body systems~\cite{BaymPethick1991book}. 

Experiments have been conducted with impurity atoms immersed in Fermi gases, forming Fermi polarons~\cite{Schirotzek2009,Kohstall2012,Scazza2017}, as well as in BECs, where they form Bose polarons~\cite{jorgensen2016,hu2016,Yan2020}. Spectroscopic as well as interferometric work have explored the properties of Bose polarons as a function of interaction strength for both attractive and repulsive impurity-medium interaction~\cite{ardila2019,Yan2020,Skou2020,SkouPRR2022,Etrych2024,Morgen2025QBeat}.

A simple variational wave function truncated at 2-body correlations, the so-called Chevy ansatz \cite{Chevy2006}, between the impurity and the environment turns out to describe the Fermi polaron remarkably well even for strong interactions~\cite{Massignan2014}. This is due to the Fermi exclusion principle which surpresses higher-order correlations for a short-range interaction~\cite{Combescot2007}. A similar wave function also agrees reasonably well with experimental data for the Bose polaron, however with considerable discrepancies for strong interactions. Moreover, experiments have so far observed a significant broadening of the spectrum for strong impurity-boson interactions. This indicates that the Bose polaron probed in experiments may not be the ground state but rather an excited state with a finite lifetime. It can nevertheless dominate the spectral response, since its wave function mainly involves 2-body correlations and, therefore, has a large overlap with the non-interacting state~\cite{massignanrev2025,Alhyder2026}. 

Here, states at lower energies due to two distinct mechanisms are explored. First, single impurities can be dressed by more than two bosonic excitations, leading either to more heavily renormalized quasiparticles~\cite{Levinsen2015,Yoshida2018} or the formation of cluster states~\cite{Christianen2024}. Second, two polarons can form a {bipolaron bound state due to an attractive interaction mediated by the surrounding BEC~\cite{Camacho-Guardian2018,Naidon2018,Jager_2022}. These low-energy states are, however, difficult to observe due to their small overlap with a non-interacting state, leading to a suppressed spectral weight.

To explore the possible existence of such low-energy states, we employ a spectroscopic pump-probe sequence, where a spectrally broad pump first creates impurities in the BEC. During an evolution time, these impurities can form low-energy states through interactions with the BEC, after which they are probed using a spectrally narrow ejection pulse. We observe a strong signal due to the formation of a Bose polaron in agreement with the Chevy ansatz. Furthermore, significant spectral weight appears at energies below the Bose polaron signal, which we explore as a function of interaction strength and evolution time. This signal is compared to the two distinct states outlined above.

\begin{figure}[t!]
    \centering
    \includegraphics[width=0.7\columnwidth]{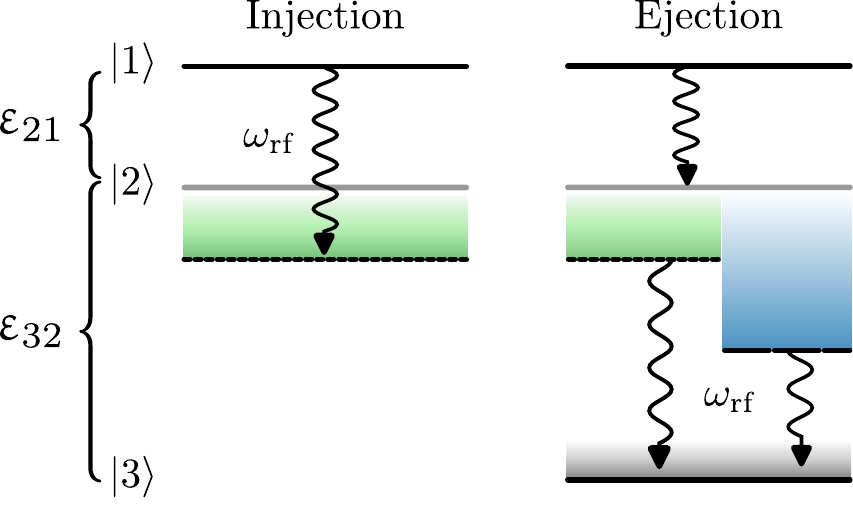}
    \includegraphics[width=\columnwidth]{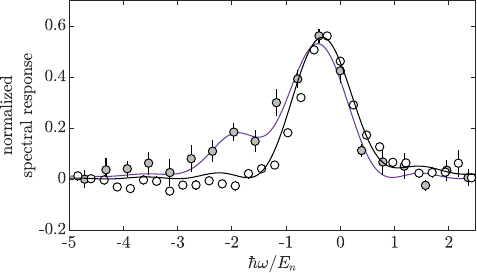}
    \caption{Experimental injection and ejection spectroscopy sequence with the relevant states. (top) In ejection measurements, an initial pump pulse populates the impurity state $\ket{2}$ from the medium state $\ket{1}$, and after a variable evolution time, a probe pulse ejects the impurity into a third state $\ket{3}$. The energy of the attractive polaron (green), low-lying impurity states (blue), and the final state (grey) with their many-body continua (shading) are illustrated. (bottom) Comparison of an injection (open circles) and ejection (filled circles) spectrum at large interaction strength $1/k_na = -0.2$. Additional spectral weight is observed at low energies for the ejection data compared to the injection data. The solid lines are fits to the data (see text).}
    \label{fig:intro}
\end{figure}

\paragraph{Experimental sequence--} The experimental ejection spectroscopy sequence is illustrated in Fig.~\ref{fig:intro}, and compared with injection spectroscopy. The starting point is a BEC of $^{39}\text{K}$ in the $\ket{F=1,m_F=-1}\equiv\ket{1}$ medium state confined in an optical dipole trap~\cite{wacker2015}. Using a very short ($\approx 1\mu$s) radio frequency (rf) pulse, a superposition with a small fraction in the $\ket{F=1,m_F=0}\equiv\ket{2}$ impurity state is created. This pump pulse is followed by an evolution time, where the impurity and medium atoms interact at a chosen interaction strength. Finally, a second tunable rf ejection pulse, in the vicinity of the $\ket{2}$ to $\ket{F=1,m_F=+1}\equiv\ket{3}$ transition, is applied. When the frequency of this probe pulse matches the energy of an impurity state, impurities are ejected to the $\ket{3}$ state. By scanning the frequency of the probe pulse, the energy of possible impurity states formed during the evolution time can be recorded. The interaction strength between impurities and medium atoms, given by the scattering length $a$, is controlled by utilizing a Feshbach resonance~\cite{lysebo2010,Tanzi2018}. In the vicinity of this resonance, the other relevant scattering lengths involving the medium state are approximately constant $a_{11}=10a_0$, $a_{13}=-33a_0$, and much smaller than the scattering lengths between the medium and impurity atoms probed here.

Due to three-body recombination between impurity and medium atoms, the lifetime of the impurities is finite~\cite{Skou2020,Morgen2025Threebody}, which prevents a direct observation of the impurities for large interaction strengths. Instead, the experimental signal is inferred from the loss of medium atoms~\cite{SupplementaryMaterial}. 

The medium density $n$ determines the characteristic energy $E_n = \hslash^2k_n^2/2m$ and wavenumber $k_n = \left(6\pi^2 n \right)^{1/3}$ of the system. However, this density decreases due to the loss processes during the evolution time and, therefore, it is evaluated at the end of the evolution time for each ejection spectrum according to Ref.~\cite{Morgen2025Threebody}.

\paragraph{Observation of low-energy impurity states--} Figure~\ref{fig:intro} shows ejection spectrum for large impurity-boson interaction strength, $1/k_na=-0.2$, for an evolution time of 30~$\mu$s, and 20~$\mu$s probe pulse duration. For comparison, an injection spectrum is shown for the same interaction strength and probe pulse duration. The ejection spectrum is given as a function of $\hslash\omega = \hslash\omega_ {\text{rf}} - \varepsilon_{32}$, while the injection spectrum is plotted as a function of $\hslash\omega = \varepsilon_{21} - \hslash\omega_{\text{rf}}$. This allows for a direct comparison of the position of the polaron peak and any low-energy signals between the two spectra. Crucially, while the two spectra agree on the position of the main polaron peak, the ejection spectrum has large spectral weight at low energies, which is not matched in injection spectroscopy. In addition, there is no significant weight at frequencies above the polaron in injection spectroscopy, as one would expect if such a signal was caused by the many-body continuum. We thus attribute the observed spectral weight to impurity states at lower energies than the polaron. 

\begin{figure*}[t!]
    \begin{center}
        \includegraphics[width = 0.8\textwidth]{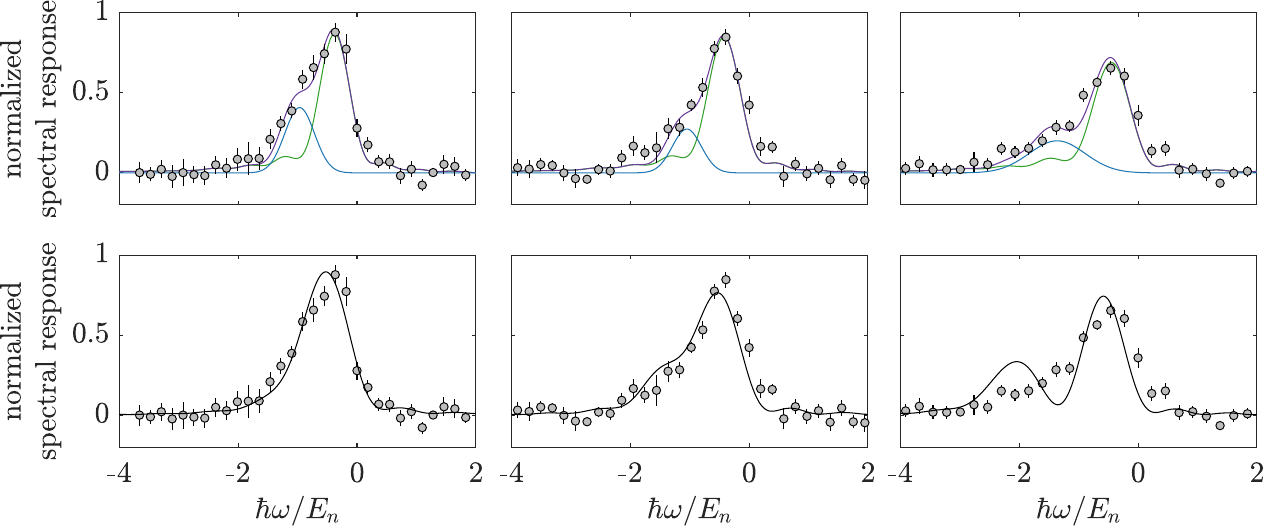}
        \caption{Ejection spectra at interaction strengths of (left) $1/k_na = -0.5$, (center) $-0.35$, and (right) $-0.2$ for 10~$\mu$s evolution time, 40~$\mu$s probe pulse duration and $20\%$ impurity concentration. (top) The prominent polaron peak is fitted with the spectral function in Eq.~\eqref{eq:A_zwierlein} (green line). The signal attributed to low-lying impurity states is fitted with a Gaussian function (blue line). The sum of both fits is also shown (purple line). (bottom) The same experimental data is compared to a combination of the spectral functions for the polaron and a zero momentum bipolaron including decay (see Eq.~\eqref{Ibip}). Only one fit parameter is used, reflecting the relative weight of the polaron and bipolaron signals.}
        \label{fig:spectra}
    \end{center}
\end{figure*}

Figure~\ref{fig:spectra} shows ejection spectra at different interaction strengths. These spectra consistently show the polaron peak, as well as significant low-energy spectral weight. The well-known attractive polaron arises from the interaction between the impurity and the surrounding medium~\cite{massignanrev2025,grusdtrev2024}, and we apply the widely used Chevy ansatz truncated at 2-body correlations to analyze this peak~\cite{Li2014}, 
which is equivalent to the ladder approximation~\cite{Rath2013}. It gives the ejection spectrum of the Bose polaron of the form~\cite{Yan2020}
\begin{equation}
    \begin{split}
        E_n \cdot A_P(\omega) = &2\pi Z_P\delta \left(\frac{\hslash \omega-E_P}{E_n}\right)+\theta \left(\frac{E_P-\hslash \omega}{E_n}\right) \\ & \times \frac{3\pi}{2\sqrt{2}}Z_P 
        \left(\frac{E_P}{\hslash \omega}\right)^2\sqrt{\frac{E_P-\hslash \omega}{E_n}},
    \end{split}
    \label{eq:A_zwierlein}
\end{equation}
where $\omega$ is the probe frequency defined above, $Z_P$ is the quasiparticle residue, $E_P$ is the polaron energy, and $\delta(x)$ and $\theta(x)$ are the Dirac delta and Heaviside functions, respectively. In addition, Eq.~\eqref{eq:A_zwierlein} assumes an ideal BEC, which is a good approximation due to the low medium scattering length $a_{11}$.

\paragraph{Extracting the energy of low-lying states--} We first fit the polaron signal in the spectra of Fig.~\ref{fig:spectra} with the spectral function Eq.~\eqref{eq:A_zwierlein} using the polaron energy $E_P$ and an overall amplitude $\mathcal{A}$ as fitting parameters. The fit is restricted to energies $\hslash\omega\geq 2 E_P$, above twice the theoretical polaron energy~\footnote{This is relaxed for data sets at low interaction strength, where the polaron signal adequately describes the whole spectrum.}, which initially excludes the low-energy spectral weight. In the fit procedure, broadening effects due to the inhomogeneous atomic density distribution and the probe pulse duration and shape are taken into account~\cite{SupplementaryMaterial}. Note that the main contribution to the width of the resulting spectral function arises from the probe pulse and results in a sinc-like shape. Based on this procedure, the fitted polaron spectral function matches the experimental data well, consistently capturing the observed main polaron peak. The extracted polaron energies are shown in Fig.~\ref{fig:LLPS_energy} and agree well with the Chevy ansatz. This agreement is consistent with previous observations of the Bose polaron and builds confidence in the application of Eq.~\eqref{eq:A_zwierlein} to model the polaron spectrum, including experimental broadening effects. Importantly, it also acts as a benchmark for the accurate determination of the density of the system, taking three-body recombination losses into account.

\begin{figure}[t!]
    \centering
    \includegraphics[width=\columnwidth]{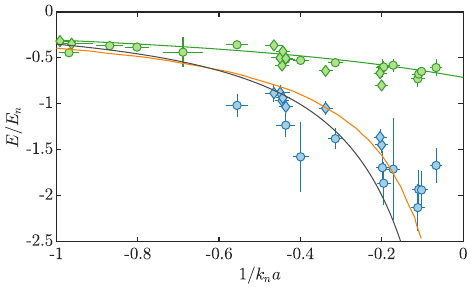}
    \caption{Extracted energies of the polaron (green) and low-lying impurity state (blue). Solid lines show the theoretical polaron energy from the Chevy ansatz (green), the energy of heavily dressed impurities from the coherent ansatz (orange), and the bipolaron ground state energy (black). Results using both 20~$\mu$s (circles) and 40~$\mu$s (diamonds) probe pulse durations are shown. Error bars are $1\sigma$ confidence intervals of the fitting procedure.}
    \label{fig:LLPS_energy}
\end{figure}
We now turn to the spectral weight observed at energies below the polaron signal.  In principle, the analysis of the low-energy signal requires theoretical knowledge of the expected spectral function~\footnote{Indeed, even the use of the expected energy dependence of an assumed low-energy state would enter the analysis due to the trap averaging process.}. To avoid any bias due to a specific spectral function, the low-energy signal is instead analyzed using a Gaussian fit function~\cite{jorgensen2016,Etrych2024}. This allows for a comparison to different theoretical predictions. The use of a Gaussian is also motivated by the fact that a large contribution to the width is due to the convolution with the probe pulse shape~\cite{SupplementaryMaterial}. 
Note, that the term in Eq.~\eqref{eq:A_zwierlein} proportional to $\theta(E_P-\omega)$ includes a spectral signal below the polaron peak due to the many-body continuum of final states, however this does not account for the large spectral weight of the observed signal in Fig.~\ref{fig:intro}.

The fit to the spectra in Fig.~\ref{fig:spectra}, thus, consists of the broadened polaron spectral function (Eq.~\eqref{eq:A_zwierlein}) together with a Gaussian function with free central energy and width. To restrict the number of fit parameters, the polaron spectral function is fixed to its previously fitted energy and only its overall amplitude, $\mathcal{A}$, is allowed to vary. The central energy of the fitted Gaussian provides the energy of the low-energy signal, which is shown as a function of the interaction strength in Fig.~\ref{fig:LLPS_energy}. These energies consistently lie below the polaron energy and decrease with increasing interaction strength. In the following, we, thus, compare the low-lying energies with two possible models of their physical origin.

\paragraph{Comparison with heavily dressed impurities--} Since the BEC is very compressible with a small boson-boson scattering length $a_{11}$, it may support the formation of deeply bound impurity states dressed by many bosonic excitations. Different theoretical approaches have been used to describe such states~\cite{massignanrev2025}, including a generalization of the Chevy ansatz to include the dressing with two~\cite{Levinsen2015} and three~\cite{Yoshida2018} excitations. Since the weak interactions allow for strong dressing, we instead choose a Gaussian ansatz, which approximately describes the correlations of the impurity with an infinite number of excitations. Gaussian ans\"atze of various degrees of sophistication have been used~\cite{Shchadilova2016,Guenther2021,Schmidt2022,Christianen2022b,Christianen2024}, and here the simplest one is chosen 
$\ket{\psi} = e^{\sum_{\mathbf{k}}( \alpha_{\mathbf{k}}\hat c^\dagger_{-\mathbf k} \hat\beta^\dagger_{\mathbf{k}} - \text{h.c.})}\ket{\rm BEC}$, 
where $\beta_{\mathbf k}^\dagger$ and $\hat c_{\mathbf k}^\dagger$ create a Bogoliubov excitation in the condensate $|\text{BEC}\rangle$ and an impurity with momentum $\mathbf k$ respectively. The variational parameters $\alpha_{\mathbf k}$ are determined by minimizing the energy~\cite{SupplementaryMaterial}. Figure~\ref{fig:LLPS_energy} shows that the ground state energy obtained from this ansatz agrees well with the experimental data. Hence, the observed low-energy signal may arise from a state where the impurity is dressed by many bosonic excitations. However, due to the dressing, this state also has a very small quasiparticle residue, strongly suppressing its spectral weight, which is instead dominated by a many-body continuum of final states~\cite{SupplementaryMaterial}. We thus expect that the spectral weight of this state is too low to provide the observed ejection spectral response. 

\paragraph{Comparison to bipolarons--} A second possible origin of the low-energy signal is the formation of bound states between two polarons during the evolution time. Such bipolarons are predicted to form \cite{Camacho-Guardian2018,Naidon2018} due to the attractive interactions between two polarons via the exchange of density modulations of the surrounding BEC~\cite{Camacho-Guardian2018b,Paredes2024}. We, therefore, calculate the binding energy $E_{BP}$ of the bipolaron relative to that of two unbound polarons $2E_P$ using an effective Schr\"odinger equation~\cite{Camacho-Guardian2018,SupplementaryMaterial}, such that the total energy of the bipolaron is $2E_P+E_{BP}$. The ejection probe pulse splits these bipolarons, and leaves one impurity behind, which forms a polaron with an energy $E_P$. Energy conservation then gives $\hslash\omega+E_P=E_{BP}+2E_P$, and therefore Fig.~\ref{fig:LLPS_energy} shows the theoretical prediction $E_{BP}+E_P$ for the bipolaron state. This is also in good agreement with experimental data, and, thus, the energy of the low-lying state is equally consistent with the formation of bipolarons. 

To explore the bipolaron interpretation further, Fig.~\ref{fig:spectra} also shows a theoretically calculated spectral function of the form $A_{\rm tot}(\omega)= r_P A_{P}(\omega)+ (1-r_P)A_{BP}(\omega)$, where $A_P(\omega)$ is the ejection spectrum of the polaron given by Eq.~\eqref{eq:A_zwierlein}, and $A_{BP}(\omega)$ is the ejection spectrum of the bipolaron given by 
\begin{equation}
\label{Ibip}
A_{BP}(\omega)=\sum_{\mathbf p} \frac{2\pi\gamma_{BP}|\psi_{BP}(\mathbf p)|^2}{\left(\hbar\omega-E_{BP}-E_P+\frac{p^2}{m}\right)^2+\gamma_{BP}^2}.  
\end{equation}
This corresponds to a zero momentum bipolaron with binding energy $E_{BP}$ and wave function $\psi_{BP}(\mathbf p)$ being split by the probe pulse into a polaron with momentum ${\mathbf p}$ and a $|3\rangle$ atom with momentum $-{\mathbf p}$ such that they both have kinetic energies $p^2/2m$. We calculate the bipolaron wave function from the effective Schr\"odinger equation. An imaginary part $\gamma_{BP}=2\gamma_c$ is added to describe bipolaron decay, where $\gamma_c$ is the damping rate of the impurity state~\cite{SupplementaryMaterial}. A broadening effect was included by convolving the theoretical result with the probe pulse shape. Finally, $r_P$ is a fit parameter, which reflects the relative weight of the polaron and bipolaron state when the probe pulse is applied. This specific fit also agrees qualitatively with the data, while differences arise at large interaction strengths. 
\begin{figure}[h]
    \includegraphics[width=\columnwidth]{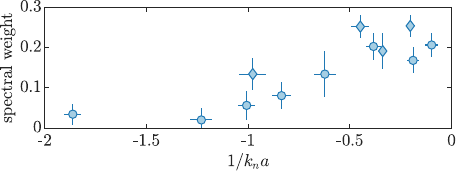}
    \includegraphics[width=\columnwidth]{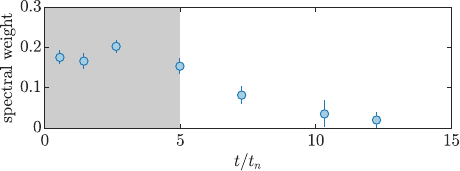}
    \caption{Spectral weight of the low-energy impurity signal for (circle) 20~$\mu$s and (diamond) 40~$\mu$s probe pulse duration. (top) Low-energy spectral weight as a function of inverse interaction strength for 20~$\mu$s and 10~$\mu$s evolution time for short and long pulse durations, respectively. (bottom) Low-energy spectral weight as a function of evolution time in units of $t_n$ at an inverse interaction strength of $1/k_na=-0.4$. The shaded region indicates evolution times shorter than the probe pulse duration. Error bars are $1\sigma$ confidence intervals from the fitting procedure.}
    \label{fig:spectralweight}
\end{figure}
\paragraph{Weight and lifetime of low-lying states--} The low-energy signal is further characterized by analyzing its spectral weight as a function of interaction strength and evolution time, as shown in Fig.~\ref{fig:spectralweight}. The spectral weight is obtained from numerical integration after subtraction of the polaron spectral function. This allows for the extraction of the small spectral weight at weak interactions $1/|k_na|>0.6$, where fits are unreliable~\footnote{Data points at closely lying interaction strengths were averaged for clarity.}. Since the total spectral weight is normalized, the low-energy spectral weight corresponds to its fractional contribution to the total spectral weight. 

Figure~\ref{fig:spectralweight} shows the low-energy spectral weight as a function of the inverse interaction strength. The signal appears at $1/k_na \approx -1$, increases in the strongly interacting regime, and saturates around $0.2$ near unitarity. The maximal observed spectral weight reflects a balance of the finite lifetime of the low-lying state and the combination of the evolution time and probe pulse duration. Importantly, the strength of the mediated interactions between impurities is expected to increase as a function of interaction strength, leading to an increase in spectral weight. This favors the interpretation of the low-energy state as a bipolaron over an impurity dressed by many excitations, which has decreasing spectral weight for strong interactions.

Figure~\ref{fig:spectralweight} shows the low-energy spectral weight as a function of the evolution time, during which the impurity and medium atoms interact. The evolution time is varied for an inverse interaction strength of $1/k_na = -0.4$, with the characteristic time of the system $t_n=\hslash/E_n\approx 4 \mu$s. At short evolution times, a close to constant spectral weight is observed, followed by a decay for longer times. We do not observe a characteristic rise time of the low-energy signal, likely because the probe pulse duration is longer than the initial formation dynamics. This long probe pulse duration is, however, necessary to spectrally resolve the low-energy signal. The decay of the signal shows that the low-energy state is short-lived compared to the polaron state, which does not show significant decay over the evolution times investigated here. Both of these results lack theoretical descriptions and thus provide a benchmark for future work.

In conclusion, we have created strongly interacting impurities in a $^{39}\text{K}$~BEC and investigated their properties using a pump-probe ejection spectroscopy sequence. The spectroscopic signal consists of a feature corresponding to the Bose polaron and significant spectral weight at lower energies. The energy and spectral weight of this low-energy signal were extracted as a function of interaction strength and evolution time. These results were compared to two theoretical models, an impurity state dressed by many bosonic excitations and a bipolaron. Both theories predict energies in agreement with the low-energy signal, while the bipolaron model is more consistent with the observed spectral weight.

Our results, thus, present a challenge for future experimental and theoretical efforts. In particular, a systematic study of impurity concentration may help distinguish between theoretical predictions. Moreover, extending such investigations to homogeneous trapping potentials or mass-imbalanced mixtures may allow for disentangling the roles of mediated interactions and few-body effects.

\begin{acknowledgments}
The authors thank Thomas Pohl for valuable discussions. We acknowledge support from the Danish National Research Foundation through the Center of Excellence "CCQ" (DNRF152) and from the Novo Nordisk Foundation NERD grant (Grant No. NNF22OC0075986). G.M.B.\ acknowledges support from the Novo Nordisk Foundation (Grant no.\ NNF23OC0086599).
K.K.N. acknowledges support from the Carlsberg Foundation through a Carlsberg Reintegration Fellowship (Grant no. CF24-1214). A.C-G. acknowledges financial support from UNAM DGAPA PAPIIT Grant No. IA101325, Project CONAHCYT No. CBF2023-2024-1765 and PIIF25.
\end{acknowledgments}

\bibliography{analysis}

\newpage

\onecolumngrid

\newpage

\section{Supplementary Material}
\subsection{Experimental signal extraction}
The spectroscopic signal presented in the main manuscript is derived from the observed loss of atoms in the system after applying the ejection sequence, which is illustrated in Fig.~1 in the main manuscript. The loss of atoms in the system is due to three-body recombination and two-body spin exchange collisions. Three-body recombination initially occurs between atoms in the medium state, $\ket{F=1,m_F=-1} \equiv \ket{1}$, and atoms in the impurity state, $\ket{F=1,m_F=0} \equiv \ket{2}$. Additionally, if the impurity atoms are transferred to the third state, $\ket{F=1,m_F=1} \equiv \ket{3}$, fast two-body spin exchange collisions between these and the medium atoms are the main loss mechanism.
\begin{figure}[h]
    \centering
    \includegraphics[width=\textwidth]{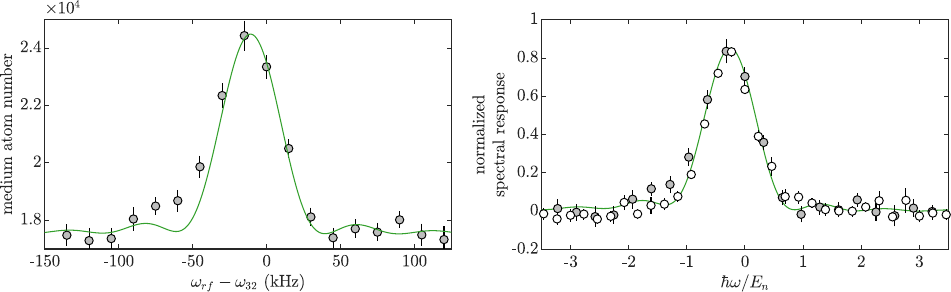}
    \caption{(left) Ejection spectroscopy data set at low interaction strength of $1/k_na = -1$, in terms of the observed medium atom number as a function of the frequency detuning of the probe pulse from the $\ket{2}$ to $\ket{3}$ unperturbed state transition. (right) Comparison between ejection (gray circles) and injection (white circles) at an interaction strength of $1/k_na = -1$ and probe pulse duration of 20 $\mu$s. The fit of the polaron spectral function to the ejection data is also shown (green line).}
    \label{fig:figure8}
\end{figure}
Ideally, one would like to measure the outcome of an ejection sequence through direct imaging of the atoms in the $\ket{2}$ or $\ket{3}$ state. However, both loss channels in the system are fast and always lead to a total loss of impurity atoms before direct detection is possible. Instead, the number of medium atoms is used as an observable, since their loss depends on the number of impurity atoms remaining in the $\ket{2}$ state or transferred to the $\ket{3}$ state at the end of the ejection sequence. In the former case, three-body recombination leads to the loss of two medium atoms per impurity atom, while in the latter, two-body spin exchange leads to only one medium atom lost per impurity atom. Thus, if the probe pulse is resonant with the $\ket{2}$ to $\ket{3}$ transition, an increased medium atom number is observed. In Fig.~\ref{fig:figure8}~(left) the observed atom number in the $\ket{1}$ medium state is shown as a function of the detuning of the probe pulse from the $\ket{2}$ to $\ket{3}$ transition, $\omega = \omega_\text{rf}-\omega_{32}$. A peak in the medium atom number is observed at a negative detuning due to the energy shift of the impurity state. Note that the maximal observed signal is proportional to the initial impurity admixture. Impurity concentrations of $\approx 20$\% are used, striking a balance between sufficient signal and remaining in the impurity regime. The lower limit is qualitatively found to be $\approx 15$\% impurity concentration for our system. The observed medium atom number constitutes the spectroscopic signal, which is normalized and scaled as follows.

First, the observed offset in the medium atom number is subtracted. The frequency axis is converted to energy, $\hslash \omega = \hslash(\omega_\text{rf}-\omega_{32})$, and scaled by the characteristic energy of the system, $E_n = \frac{\hslash^2k_n^2}{2m}$, with $k_n = (6\pi^2n)^{(1/3)}$ where $n$ is the medium density at the time of probing the system~\cite{Morgen2025Threebody}. Finally, the area under the data points is numerically normalized to one.
The resulting normalized spectroscopic signal is shown in Fig.~\ref{fig:figure8}~(right), where it is compared to an injection measurement at $1/k_na = -1$. Both the ejection and injection measurements are in excellent agreement. Note that at this interaction strength, any low-lying impurity state is expected to have an energy shift similar to the polaron. Thus, it is typically not resolved when taking experimental broadening effects into account. Hence, injection and ejection spectra overlap significantly at interaction strengths below those presented in the main manuscript.

\subsection{Experimental broadening effects}
The experimental broadening effects that affect the spectral response of the impurity are discussed in the following. The two main contributions are the inhomogeneous density distribution of the system, and the finite duration of the probe pulse.
\begin{figure}[h]
    \centering
    \includegraphics[width=\textwidth]{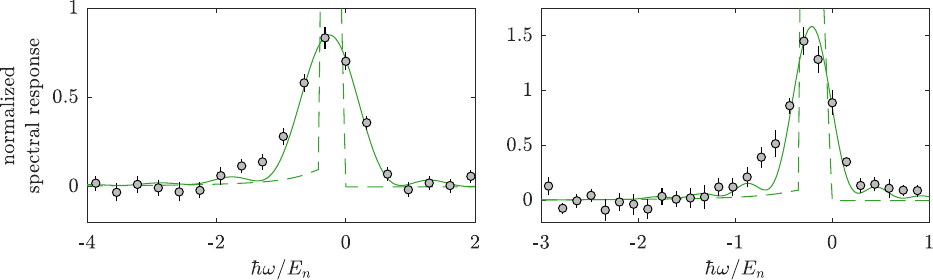}
    \caption{Ejection spectra at $1/k_na = -1$, with probe pulse durations of 20~$\mu$s (left) and 40~$\mu$s (right). The trap averaged polaron spectral function (dashed line) includes both the polaron peak and continuum. The polaron spectral function after both trap averaging and convolution with the probe pulse is also shown (solid line).}
    \label{fig:figure7}
\end{figure}
The inhomogeneous density distribution is a result of the optical dipole trap, which leads to a Thomas-Fermi distribution of the medium condensate. Thus, impurities at the center of the trap experience a higher medium density relative to impurities closer to the edge of the distribution. The impurities throughout the medium are, therefore, probed at different interaction strengths, and the observed frequency spectrum is broadened. This broadening is accounted for in the fitting procedure by numerically averaging the polaron spectral function (Eq.~1 in the main manuscript) over the condensate density distribution. The effect of this procedure is shown in Fig.~\ref{fig:figure7} for an inverse interaction strength of $1/k_na = -1$.

Secondly, the finite duration of the probe pulse leads to a broadened frequency spectrum around the carrier frequency. Thus, impurities at different energies are probed simultaneously, which also leads to a broadening of the spectral response. Since the probe pulse is a square pulse in time, it has the shape of a sinc function in frequency space. To account for this, the trap-averaged polaron spectral function is convolved with the appropriate sinc function. The result is also shown in Fig.~\ref{fig:figure7}, where the broadening due to 20~$\mu$s and 40~$\mu$s probe pulse durations is compared.

Generally, good quantitative agreement between the experimental data and the polaron spectral function is observed for low interaction strengths after taking both broadening effects into account. In particular, for 20~$\mu s$ probe pulse duration, the main contribution to the observed width is due to the pulse duration, while the broadening is reduced for 40~$\mu$s probe pulse durations, as expected.

\subsection{Spectra for different ejection pulse durations}
Here, additional ejection spectra for both 20~$\mu$s and 40~$\mu$s probe pulse duration for various impurity-medium interaction strengths are presented. These data sets have 20~$\mu$s and 10~$\mu$s evolution times for the short and long probe pulse durations, respectively, at 20\% impurity concentration. The use of 20~$\mu$s versus 40~$\mu$s probe pulse durations corresponds to two different trade-offs between a well-defined evolution time and sufficient frequency resolution of the ejection pulse.

A pulse duration of 20~$\mu$s is typically shorter or equal to the evolution time but has limited frequency resolution. On the other hand, a pulse duration of 40~$\mu$s is longer than typical evolution times, but provides better frequency resolution.

For low interaction strengths, the low-lying impurity signal may overlap significantly with the dominant polaron signal due to the aforementioned broadening effects. Therefore, the dual fit (see main manuscript) is only performed if the expected energy difference between the polaron and low-lying impurity state can be resolved with the probe pulse duration.
\begin{figure}[h]
    \centering
    \includegraphics[width=\textwidth]{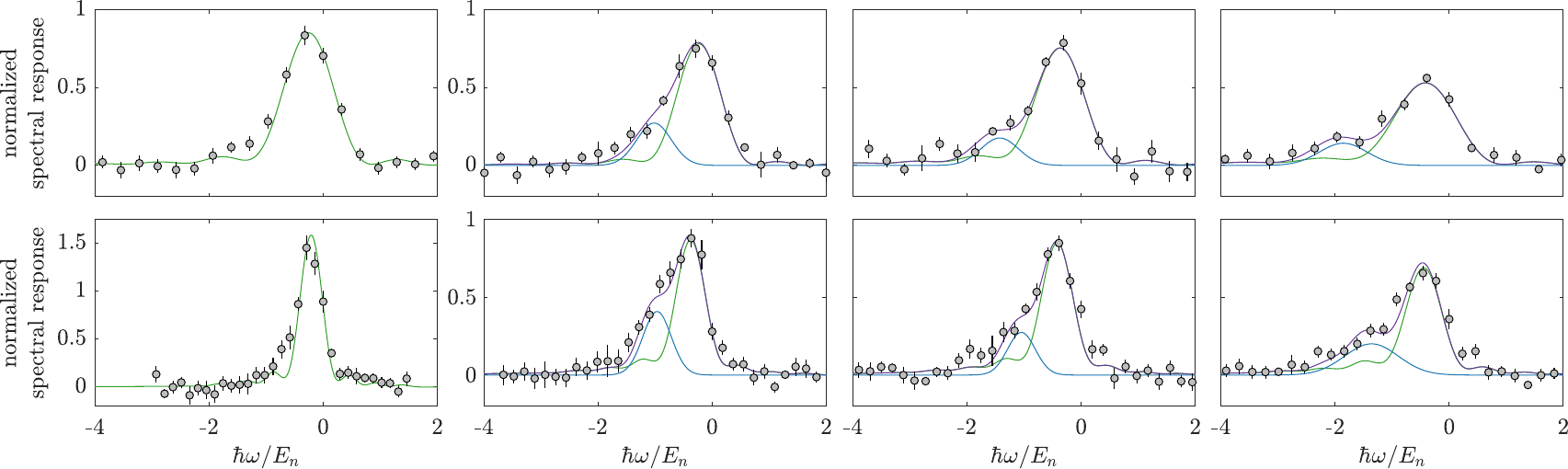}
    \caption{Ejection spectra at interaction strengths of (left-right) $1/k_na = -1, -0.5, -0.35$, and $-0.2$, with probe pulse durations of 20~$\mu$s (top row) and 40~$\mu$s (bottom row). Evolution times of 20~$\mu$s and 10~$\mu$s for the short and long probe pulse durations and 20\% impurity concentration was used. The main peak is fitted with the spectral function of the polaron from Eq.~1 in the main manuscript (green line). The low-lying impurity state is fitted with a Gaussian function (blue line), with the sum of both fits also shown (purple line).}
    \label{fig:signals_fullset}
\end{figure}
In Fig.~\ref{fig:signals_fullset} this is not the case for an interaction strength of $1/k_na = -1$, and the spectrum is only fitted using the polaron spectral function from Eq.~1 in the main manuscript. For higher interaction strengths, the energy shift is larger, and the low-lying impurity state is resolved better. In this case, the low-lying impurity signal is fitted with a Gaussian function, and the energy is extracted. The extracted energies of the polaron and low-lying impurity state are then compared to theoretical predictions, as shown in Fig.~3 in the main manuscript.

\newpage

\subsection{Theoretical Analysis}
Here, we summarize the theoretical framework employed to analyze the experimental spectra for both the polaron and bipolaron assumptions. Comprehensive discussions can be found in recent reviews~\cite{massignanrev2025,grusdtrev2024}, and earlier studies~\cite{Rath2013,Shchadilova2016,Chevy2006}. We do not enter into the details of these theories and instead focus on the interpretation of the data. 

Within Bogoliubov theory, the microscopic Hamiltonian reads
\begin{gather}
\hat H=\sum_{\mathbf p}E_{\mathbf p}\hat \beta^\dagger_{\mathbf p} \hat \beta_{\mathbf p}+\sum_{\mathbf p }(\epsilon_{\mathbf p}+n_0g)\hat c^\dagger_{\mathbf p} \hat c_{\mathbf p}+\sum_{\mathbf p,\mathbf q}\frac{g}{\sqrt{V}}\sqrt{\frac{n_0\epsilon_{\mathbf q}}{E_{\mathbf q}}}\hat c_{\mathbf p+\mathbf q}^\dagger(\hat \beta_{\mathbf q}+\hat \beta^\dagger_{-\mathbf q})\hat c_{\mathbf p}+\frac{g}{V}\sum_{\mathbf p,\mathbf p',\mathbf q}\hat b_{\mathbf p+\mathbf q}^\dagger \hat c_{\mathbf p'-\mathbf q}^\dagger\hat c_{\mathbf p'} \hat b_{\mathbf p} \end{gather}
where $\hat \beta^{\dagger}_{\mathbf p}=(u_{\mathbf p}\hat b_{\mathbf p}^\dagger+v_{\mathbf p}\hat b_{-\mathbf p})$ creates a  Bogoliubov mode with momentum $\mathbf p$ and energy 
$$E_{\mathbf p}=\sqrt{\epsilon_{\mathbf p}(\epsilon_{\mathbf p}+2n_0g_{bb})},$$ 
with $g_{bb}=4\pi a_{11}/m$ and $\epsilon_{\mathbf p}=p^2/2m.$ The Bogoliubov quasiparticles are expressed  in terms of the bare $\hat b_{\mathbf p}^\dagger/\hat b_{\mathbf p}$ creation/annihilation bosonic operators.  Here, we have the coherence factors $$u_{\mathbf p}^2=1+v_{\mathbf p}^2=\frac{p^2/2m+n_0g_{bb}}{2E_{\mathbf p}}+\frac{1}{2}.$$ The operator $\hat c_{\mathbf p}^\dagger$ creates an impurity with momentum $\mathbf p$. Finally, the impurity-boson coupling constant is $g=4\pi a/m.$ 

\subsection{Single Polaron} 
We employ two complementary theories to analyze the spectra from a single polaron perspective: a Chevy ansatz adapted to the Bose polaron and a coherent variational ansatz~\cite{Shchadilova2016}. The former is restricted to a single Bogoliubov excitation and is equivalent to the so-called non-self consistent T-matrix approximation. In contrast, the coherent ansatz incorporates dressing by, in principle, an infinite number of Bogoliubov excitations outside the condensate.

The corresponding variational states are given by~\cite{Chevy2006,Shchadilova2016}
\begin{equation}
|\psi_P\rangle =
\begin{cases}
\displaystyle
\left(
\sqrt{Z_P}\,\hat c_{\mathbf 0}^\dagger
+\sum_{\mathbf p}\psi_P(\mathbf p)\,
\hat c^\dagger_{\mathbf p}\hat \beta^\dagger_{-\mathbf p}
\right)|\mathrm{BEC}\rangle,
& \text{(Chevy ansatz)},
\\[10pt]
\displaystyle
\exp\!\left[
\sum_{\mathbf{k}}
\left(
\alpha_{\mathbf{k}}\hat c^\dagger_{-\mathbf k}\hat\beta^\dagger_{\mathbf{k}}
-\mathrm{h.c.}
\right)
\right]
|\mathrm{BEC}\rangle,
& \text{(coherent variational ansatz)}.
\end{cases}
\end{equation}

Minimization of the energy yields the polaron energy
\begin{equation}
    \frac{E_P}{E_n} =
\begin{cases}
\frac{4}{3\pi} \left[{(k_n a)^{-1} - \Xi\left(E_P/E_n,\, k_n \xi\right)}\right]^{-1}, & \text{(T-matrix / Chevy ansatz)}
   \\[10pt]\left[{(k_n a)^{-1} - (k_n a_0)^{-1}}\right]^{-1}, & \text{(coherent variational ansatz)}
\end{cases}
\end{equation}
with $\Xi(x, y)=[1-xy^2\text{atanh}(\sqrt{1+xy^2})/(\sqrt{2(1+xy^2})y\pi)],$ $\xi=1/\sqrt{8\pi a_{11}n}$ the coherence length of the BEC ~\cite{massignanrev2025}. 
The effective length $a_0$ in the coherent ansatz is given by
$$a_{0}^{-1} = \frac{4\pi}{m} 
\sum_{\mathbf{k}} \left( \frac{m}{k^{2}} 
- \frac{\epsilon_{\mathbf{k}}/E_{\mathbf k}}{E_{\mathbf{k}} + \frac{k^{2}}{2m}} \right).
$$ 
For the Chevy ansatz we obtain a self-consistent equation that can be solved numerically and is illustrated by the green line in Fig.~3 of the main text. On the other hand, the coherent ansatz yields an analytical expression for the energy, orange line in Fig.~3 of the main text

The spectral function for Chevy's ansatz acquires the closed expression in Eq.~1 (main text) for an ideal BEC. Within this approach, the polaron branch retains significant weight even at unitary, where the residue of this branch is $Z=2/3.$ 

The coherent ansatz predicts a much deeper energy, since the impurity can be dressed by many Bogoliubov excitations. This results in a strong suppression of the quasiparticle residue, which essentially vanishes for strong interactions. Indeed, the polaron branch becomes essentially invisible in this regime.

A direct comparison of the predictions for the energy of the T-matrix and the coherent variational ansatz is shown in Fig.~3 (main text). Further comparison of the distribution of the spectral function can be found in the Supplementary Material of Ref.~\cite{Shchadilova2016}, where it is made explicit that the T-matrix predicts a polaron state with significant spectral weight, whereas the coherent ansatz leads to a vanishing residue for this polaron branch.

\subsection{Bipolarons} 
The medium can induce an interaction between two polarons which can be strong enough to support bound-states i.e., bipolarons~\cite{Camacho-Guardian2018,Naidon2018,massignanrev2025,grusdtrev2024}. In the static approximation for the mediated interaction, bipolarons can be described from an effective Schrödinger equation for the relative motion of two polarons~\cite{Camacho-Guardian2018},
\begin{equation}
\varepsilon_{\mathrm{BP}}\,\psi_{BP}(\mathbf p) = 2E_P({\mathbf p})\psi_{BP}(\mathbf p)
+ \sum_{k'} V_{\mathrm{eff}}(\mathbf p,\mathbf k') \psi_{BP}(\mathbf k').
\end{equation}
This equation yields the bipolaron binding energy $E_{\mathrm{BP}},$ where $E_P({\mathbf p})$ is the single polaron dispersion $\mathbf p.$ We approximate $E_P({\mathbf p})=E_P+p^2/2m$ where $E_P$ is the zero-momentum polaron energy. The bipolaron energy is $\varepsilon_{\text{BP}}=E_{\text{BP}}+2E_P,$ that is, $E_{\text{BP}}$ is the binding energy of the bipolaron shown in the main text.
Here, 
$$V_{\text{eff}}(\mathbf p,\mathbf p')=nZ({\mathbf p})Z({\mathbf p'})\left[2\mathcal T(\mathbf p,E_P)\mathcal T(\mathbf p',E_P)G_{11}(\mathbf p-\mathbf p',0)+\mathcal T^2(\mathbf p,E_P)G_{12}(\mathbf p-\mathbf p',0)+\mathcal T^2(\mathbf p',E_P)G_{12}(\mathbf p-\mathbf p',0)\right],$$ 
where $\mathcal T(\mathbf p,\omega)$ is the $T$-matrix for the impurity-boson scattering~\cite{Rath2013}, here the quasiparticle residue  of the single polaron $Z({\mathbf p})$ renormalizes the polaron-polaron interactions. Finally, we have $G_{11}(\mathbf p,\omega)=u_{\mathbf p}^2/(\omega-E_{\mathbf p})-v_{\mathbf p}^2/(\omega+E_{\mathbf p})$ the normal and $G_{12}(\mathbf p,\omega)=u_{\mathbf p}v_{\mathbf p}/(\omega-E_{\mathbf p})-u_{\mathbf p}v_{\mathbf p}/(\omega+E_{\mathbf p})$ the anomalous BEC Green's function.

The bipolaron state can be written as $|\text{BP}\rangle=\sum_{\mathbf p}\psi_{BP}(\mathbf p)\hat a^\dagger_{\mathbf p}\hat a^\dagger_{-\mathbf p}|\text{BEC}\rangle,$ where $\hat a_{\mathbf p}$ denotes the annihilation operator of a polaron with momentum $\mathbf p$. The bipolaron contribution to the spectral response is given by
\begin{gather}
A_{\text{BP}}(\omega)=\sum_{\mathbf p}|\langle f_{\mathbf p}|\hat H_{rf}|\text{BP}\rangle|^2\delta\left(-\hbar\omega-[E_f(\mathbf p)-\varepsilon_{BP}]\right).
\end{gather}
Here, the initial state is formed by the bipolaron, while the final state after ejection consists of a polaron plus one free atom transferred into the third state. The final energy is $E_f(\mathbf p)=E_P+p^2/2m+{p^2/2m},$ such that conservation of energy reads as $-\hbar\omega = E_f(\mathbf p) - \varepsilon_{\mathrm{BP}} = \frac{p^2}{m} + E_P - E_{\mathrm{BP}}.$ 

Taking $\hat H_{\text{rf}}\propto \sum_{\mathbf p}\hat c^{\dagger}_{3,\mathbf p} \hat c_{2,\mathbf p}+\text{h.c}$ and including finite lifetime effects (we replace the Dirac delta by a Lorentzian), the dominant contribution to the bipolaron spectral function is given by 
\begin{gather}
A_{\text{BP}}(\omega)\propto \sum_{\mathbf p}|\psi_{\text{BP}}(\mathbf p)|^2\frac{\gamma_{\text{BP}}}{(\hbar\omega+p^2/m-E_{\text{BP}}-E_P)^2+\gamma_{\text{BP}}^2}.
\end{gather}

We use an experimentally obtained damping rate for the impurity state of $2\gamma_c/E_n=0.0513, 0.0716, \text{and}\, 0.0812$ for $1/k_na=-0.5, -0.35,\text{and} -0.20,$ respectively~\cite{Morgen2025Threebody}, and assume that $\gamma_{\text{BP}}=2\gamma_c$. In our numerics, both the single polaron spectral function in Eq.~1 (main text) and the bipolaron functions are normalized such that $\frac{1}{2\pi}\int d\omega A_{P}(\omega)=\frac{1}{2\pi}\int d\omega A_{\text{BP}}(\omega)=1.$ The total spectral response is then modeled as
\begin{equation}
A(\omega)
=
\frac{1}{2\pi}
\left[
r_P A_P(\omega)
+
(1 - r_P) A_{\mathrm{BP}}(\omega)
\right],
\end{equation}
where $r_P$ is a fitting parameter that controls the relative weight of the polaron and bipolaron contributions to the spectral signal.

We include an additional broadening due to the finite duration of the probe pulse. The total spectral function is then given by
\begin{gather}
I_{\text{tot}}(\omega)
=
\frac{1}{\mathcal N}
\int d\omega' \;
A(\omega') \,
\mathcal F(\omega - \omega'),
\end{gather}
with $\mathcal N$ a normalization factor and $\mathcal F(\omega)$ the frequency response. For a probe pulse of duration $\tau$ the  frequency response is given by
\begin{gather}
\mathcal F(\omega)
=
\left|
\frac{\sin(\omega \tau / 2)}{\omega\tau/2}
\right|^2 .
\end{gather}
Thus, our theoretical analysis includes the polaron and bipolaron functions, the inherent decay rate of the impurity state, and the broadening induced by the finite duration of the probe pulse.
\end{document}